\documentclass{ws-ijmpd}

\begin{document}

\markboth{M. Perucho, V. Bosch-Ramon and D. Khangulyan} {On the interaction
of jets with stellar winds in XRBs}

%
\catchline{}{}{}{}{}
%

\title{On the interaction of jets with stellar winds in massive X-ray binaries.}

\author{Manel Perucho}

\address{Departament d'Astronomia i Astrof\'{\i}sica, Universitat de Val\`encia, C/ Dr. Moliner 50, 46100, Burjassot, Valencian Country, Spain.\\
manel.perucho@uv.es}

\author{Valent\'{\i} Bosch-Ramon, Dmitry Khangulyan}

\address{Max-Planck-Institut f\"ur Kernphysik, Saupfercheckweg 1, Heidelberg 69117, Germany.\\
vbosch@mpi-hd.mpg.de, Dmitry.Khangulyan@mpi-hd.mpg.de}

\maketitle
\begin{history}
\received{15 December 2009}
\revised{2 March 2010}
\comby{Managing Editor}
\end{history}

\begin{abstract}
We present the first three-dimensional simulations of the evolution of a microquasar jet inside the binary-star system. The aim is to study the interaction of these jets with the stellar wind from a massive companion and the possible locations of high-energy emission sites. We have simulated two jets with different injection power in order to give a hint on the minimum power required for the jet to escape the system and become visible in larger scales. In the setup, we include a massive star wind filling the grid through which the jet evolves. We show that jets should have powers of the order of $10^{37}\rm{erg/s}$ or more in order not to be destroyed by the stellar wind. The jet-wind interaction results in regions in which high energy emission could be produced. These results imply the possible existence of a population of X-ray binaries not detected in the radio band due to jet disruption inside the region dominated by the stellar wind. 
\end{abstract}


\section{Introduction}
The powerful jets of X-ray binaries (XRB; i.e. microquasars) are produced
close to a compact object, a black hole or a neutron star, through ejection of material
accreted from the companion star. The interaction
of jets with the stellar wind in the binary region has to be taken
into account as a possible source of strong shocks and/or jet
disruption in the case of a massive companion star. The occurrence
of collisionless shocks can lead to efficient particle acceleration
(see Ref.~\refcite{ri07}), resulting in significant non-thermal
emission of synchrotron and inverse Compton origin and, possibly,
from proton-proton collisions (see Ref.~\refcite{brk09} and
references therein).

In Ref.~\refcite{pbr08} (PBR08 from now on) the authors performed
two-dimensional (2D) numerical simulations in order to show how the
strong wind of an OB star can influence the jet dynamics at scales
similar to the orbital separation ($\sim 0.2$~AU). Simulations in
two dimensions of a hydrodynamical jet interacting with an
homogeneous (i.e. not clumpy) stellar wind were performed in
cylindrical (axisymetric) and planar (slab) symmetry. The results
showed that a strong recollimation shock is likely to occur at jet
heights $\sim 10^{12}$~cm, which could lead to efficient particle
acceleration and explain the TeV emission seen in several high-mass
X-ray binaries (LS~5039, LS~I~+61~303, Cygnus X-1, see Refs.~\refcite{ah05},
\refcite{al06} and \refcite{al07}). It was also found that jet disruption
could occur for jet kinetic luminosities as high as $L_{\rm j}\sim
10^{36}$~erg~s$^{-1}$ because of jet instabilities produced by the
act of the strong and assymetric wind. Such a $L_{\rm j}$-value is $\sim
0.1-1$\% the Eddington luminosity, typical for an X-ray binary with
persistent jets (see Fig.~1 in Ref.~\refcite{fe03}). This implies that
the stellar wind in high mass microquasars could play a role not only
for feeding accretion, but also for high-energy radiation and jet
suppresion at the binary system scales. To carry this research further, we have performed
three-dimensional (3D) simulations of hydrodynamical jets interacting with a
strong stellar wind. We find, confirming previous results, that even
jets with $L_{\rm j}> 10^{36}$~erg~s$^{-1}$ may be disrupted, since
the lower wind-jet momentum transfer in 3D as compared to 2D is balanced by the development
of helical Kelvin-Helmholtz (KH) instabilities. Our findings also point
to the presence of a strong recollimation shock that could
efficiently accelerate very energetic particles.

\section{Numerical Simulations}

\subsection{Set-up}
We have performed 3D simulations of two fast supersonic
hydrodynamical jets with $L_{\rm j}=10^{35}$ (Jet 1) and
$10^{37}$~erg~s$^{-1}$ (Jet 2), and injection velocity $1.66\times
10^{10}$~cm~s$^{-1}$. The medium in which the jets propagate is
an isotropic wind (as seen from the star) of mass loss rate
$10^{-6}$~$M_\odot$~$\rm{yr}^{-1}$ and
(constant) velocity $2\times 10^8$~cm~s$^{-1}$, typical for an O-type star (see Section 2 and
Table 1 in PBR08). The initial jet densities
are different, being $\rho_1=0.088\,\rho_{\rm a}$ and
$\rho_2=8.8\,\rho_{\rm a}$, where $\rho_{\rm a}=3\times
10^{-15}$~gr~cm$^{-3}$ is the stellar wind density. The initial jet
temperature is $T_{\rm j}\simeq 10^{10}\,\rm{K}$, Mach number is
$M_{\rm j}=16.6$ and pressure $P_{\rm j,1}=70.8\,\rm{dyn\,cm^{-2}}$
and $P_{\rm j,2}=7.1\times10^3\,\rm{dyn\,cm^{-2}}$. The star is
located at a distance of $2\times 10^{12}$~cm from the base of the jet, in a
direction perpendicular to the jet axis. Hereafter, we will use coordinate $z$ for the direction of propagation of the jet, $x$ for the direction connecting the jet base and the star, and $y$ for the direction perpendicular to both.

We have used a finite-difference code, named \textit{Ratpenat},
which solves the 3D equations of relativistic hydrodynamics
written in conservation form. \textit{Ratpenat} has been
parallelized using a hybrid scheme with both parallel processes
(MPI) and parallel threads (OpenMP) inside each process (see
Ref.~\refcite{pm10} for further details). The simulations have been
performed in Mare Nostrum, at the Barcelona Supercomputing Centre
(BSC) using up to 128 processors. The numerical grid box expands
transversally 20~$R_j$ (jet radii) on each side of the jet axis (making
a total of 40~$R_j$), and 10~$R_j$ per node (4 processors). The
numerical resolution of the simulation is of 4 cells per initial jet
radius. This means that the final box is $160\times160\times1280$
cells. An extended grid is used in the transversal direction, with
80 cells, which brings the outer boundary 80~$R_j$ farther from the
axis. The resolution is relatively low due to the amount of computational time
needed to perform the simulations. However, as the jets initially
expand, the effective resolution at the distances of interest, say
$\sim 10^{12}$ from the base, is $\sim 16$ cells per jet radius.

We implicitly assume that the magnetic field has no dynamical
influence in the evolution of the jet, that the wind is continuous
and homogeneous during the time of the simulations and that the
compact object is at the same orbital position during this time
($10^2-10^3\,\rm{s}$ compared to orbital times $T >
10^5\,\rm{s}$).

\subsection{Results}
\begin{figure}[t]
\centerline{\psfig{file=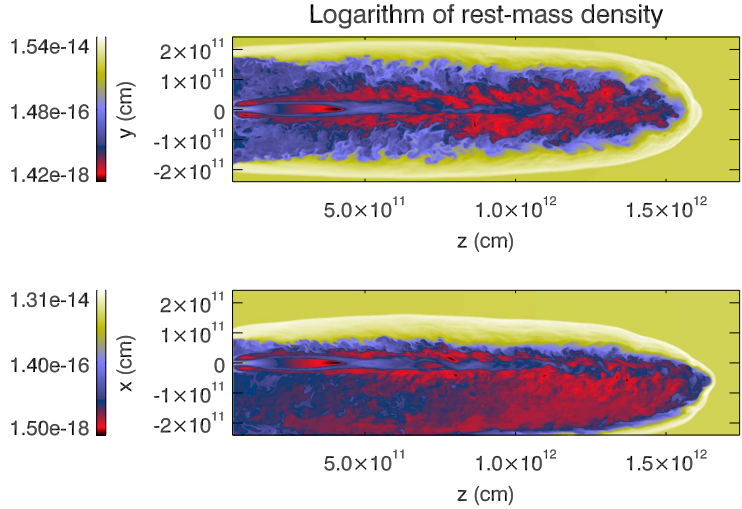,width=\textwidth}}
\vspace*{8pt}
\caption{ Jet 1, cuts of rest-mass density (in $\rm{g\,cm^{-3}}$) along the propagation axis. The upper panels show a cut perpendicular to the plane of symmetry of the wind (YZ plane), whereas the bottom panels show a cut in this plane (XZ plane). \label{fig:maps1}}
\end{figure}

Jet 1 propagates up to $z\simeq 1.6\times10^{12}\,\rm{cm}$ after
$\simeq 1.25\times10^3\,\rm{s}$. In the beginning of the simulation,
the jet expands and generates a thick shear layer with positive
velocities in the jet direction. The backflow surrounds and interacts strongly with this outer jet region,
generating instabilities that grow in the shear layer. These
instabilities are asymmetric in the plane of impact of the wind due to the different pressures in the
cocoon on both sides of the jet (see Fig.~\ref{fig:maps1}). After expansion, the central region of
the jet becomes underpressured with respect to its surroundings and
recollimates until the formation of a reconfinement shock. This quasi-steady
shock propagates very slowly from $z\simeq
2\times10^{11}\,\rm{cm}$ to $z\simeq 4\times10^{11}\,\rm{cm}$, as
the pressure in the cocoon drops (see PBR08). Downstream of the
reconfinement shock, the instabilities, which were growing in the
shear layer, propagate to the whole section of the jet as the
internal jet flow is decelerated and becomes more sensitive to
perturbations. This process ends up in the mixing and deceleration
of the jet flow at $z\geq 10^{12}\,\rm{cm}$.

\begin{figure}[h]
\centerline{\psfig{file=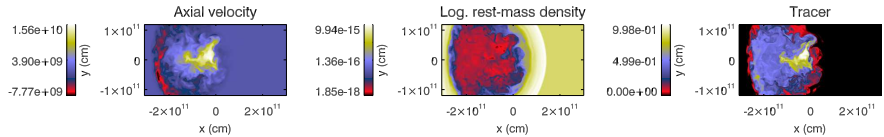,width=\textwidth}}
\vspace*{8pt}
\caption{ Jet 1, transversal cuts of axial velocity (in $\rm{cm\,s^{-1}}$), rest-mass density (in $\rm{g\,cm^{-3}}$) and jet mass fraction at $z\simeq 1.3\times10^{12}\,\rm{cm}$. \label{fig:mapst1}}
\end{figure}

Fig.~\ref{fig:maps1} shows two axial cuts of Jet 1 of rest-mass density, along
the YZ (upper) and the XZ planes (lower) at the
last snapshot. The deviation caused by the wind thrust can be
observed in the XZ plane, which is the plane of symmetry of
the wind. Fig.~\ref{fig:mapst1} shows transversal cuts of axial velocity,
logarithm of rest-mass density and tracer\footnote{The tracer,
$f=\left[0,1\right]$, indicates the composition of the fluid, with 0
corresponding to pure wind material, 1 to pure jet material and any
value between 0 and 1 indicates the relative amount of jet material
in a cell.} at $z\simeq 1.3\times10^{12}\,\rm{cm}$. In this figure,
we can see the deformation of the bow-shock caused by the wind
thrust. The jet has been entrained, at this axial position, by the
wind material --$f_{max}<1$, and the maximum velocity in the jet
fluid is still relatively fast ($\simeq
1.5\times10^{10}\,\rm{cm\,s^{-1}}$), despite the irregular
morphology and mixing. Actually, the velocity drops to $v\simeq
10^{10}\,\rm{cm\,s^{-1}}$ after this point. The low velocity of the jet at the end of the simulation,
along with its final (destabilized) structure, implies that the jet
will be disrupted and will not propagate collimated out of the binary system.

\begin{figure}[t]
\centerline{\psfig{file=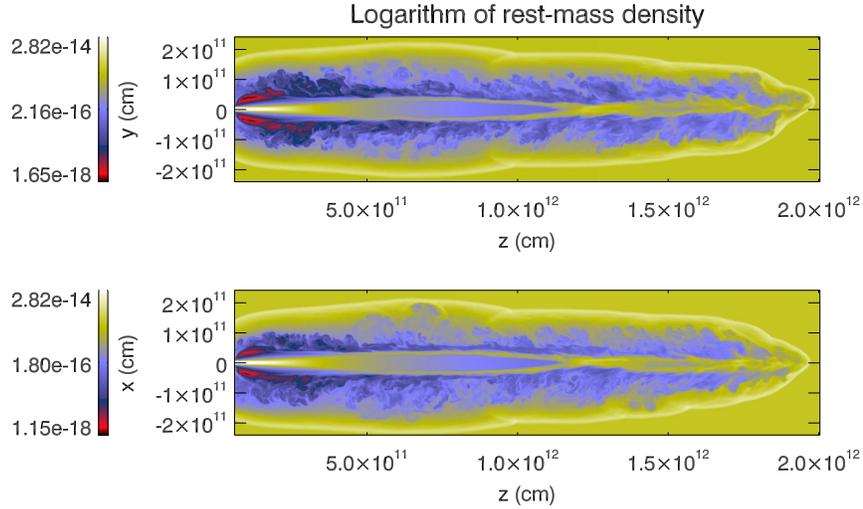,width=\textwidth}}
\vspace*{8pt}
\caption{ Jet 2, cuts of rest-mass density (in $\rm{g\,cm^{-3}}$) along the propagation axis. The upper panels show a cut perpendicular to the plane of symmetry of the wind, whereas the bottom panels show a cut in this plane. \label{fig:maps2}}
\end{figure}

\begin{figure}[h]
\centerline{\psfig{file=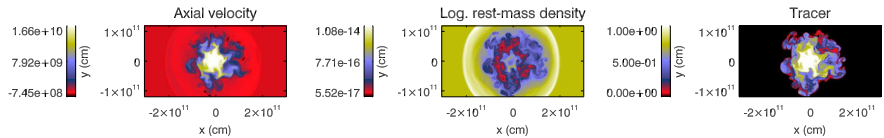,width=\textwidth}}
\vspace*{8pt}
\caption{ Jet 2, transversal cuts of axial velocity (in $\rm{cm\,s^{-1}}$), rest-mass density (in $\rm{g\,cm^{-3}}$) and jet mass fraction at $z\simeq 1.5\times10^{12}\,\rm{cm}$. The upper panels show a cut perpendicular to the plane of symmetry of the wind, whereas the bottom panels show a cut in this plane. \label{fig:mapst2}}
\end{figure}

The evolution of Jet 2 is very similar to that of Jet 1 from a
qualitative point of view, but this jet propagates up to $z\simeq
2\times10^{12}\,\rm{cm}$ in just $\simeq 210\,\rm{s}$. The jet expands
more at the base because it is initially denser (more overpressured)
and the velocity of the jet head is faster. Therefore, the
reconfinement shock is stronger and occurs farther downstream (see
PBR08). The location of the shock changes with time from $z\simeq
6\times 10^{11}\,\rm{cm}$ to $z\simeq 10^{12}\,\rm{cm}$ (see
Fig.~\ref{fig:maps2}). The effect of the wind in the direction of
the jet propagation is very small, as seen in Figs.~\ref{fig:maps2}
and \ref{fig:mapst2}. In the latter, the structure of the bow shock
at $z\simeq 1.5\times10^{12}\,\rm{cm}$ is observed to be more
symmetric than that in Jet 1 (Fig.~\ref{fig:mapst1}). The
devolopment of asymmetric KH instabilities in the shear layer, as in
Jet 1, propagates to the whole jet after the reconfinment shock, what
triggers helical motions and distortions in the jet.
Fig.~\ref{fig:mapst2} shows that the jet core is unmixed ($f=1$) and
that the flow velocity is still as high as that in the injection
point. At the end of the simulation, the velocity of the bow-shock
ahead of the jet is $\simeq 8.4\times10^{9}\,\rm{cm\,s^{-1}}$, which
is close to the initial speed ($\simeq
9\times10^{9}\,\rm{cm\,s^{-1}}$).

\section{Discussion}
Our work shows that the stellar wind from a massive star may cause
the disruption of a jet if its power is $\leq
10^{37}\,\rm{erg\,s^{-1}}$. The cocoon generated by the jet presents
asymmetric properties due to the impact of the wind. The difference in
pressure on both sides of the jet triggers helical perturbation
modes in the jet, which develop initially only in the shear layer of the
jet. After the jet core goes through a strong reconfinement shock,
the helical instability propagates to the whole jet. In PBR08, it
was shown that such a shock will occur inside the binary region if
the temperature of the jet is $T_j < 10^{14}\,\rm{K}$ when the jet is
surrounded by the cocoon, or $T_j < 10^{13}\,\rm{K}$ when the jet is in
direct contact with shocked wind. Reconfinement shocks and those
produced in the interaction with the stellar wind (bow-shock and
reverse-shock in the wind impact region) are candidate locations for
the production of high energy emission. This aspect will be treated
elsewhere (Bosch-Ramon, Khangulyan \& Perucho, in preparation).

A candidate X-ray binary where disruption of the jet could be taking
place is LS 5039 (see Ref.~\refcite{mo08}), but higher resolution observations are required
for a proper probe of the jet-wind interaction region. The presented situation could take place in several high-mass XRBs (HMXB) in the Galaxy. The luminosity function derived in Ref.~\refcite{gr03} predicts 3 HMXBs
with $L_X=10^{35}\rm{erg/s}$. Following Ref.~\refcite{fe05}, a HMXB
with a $10\,M_\odot$ black hole, could produce a jet with kinetic
power between $10^{35}$ and $10^{38}\rm{erg/s}$, which is in the
range of the simulations performed here. Although
Ref.~\refcite{gr03} does not offer any specific prediction for $L_X
\leq 10^{35}\rm{erg/s}$, extrapolating the given luminosity function, we
deduce that there is room for a few ($\sim 10$) sources in our
Galaxy in which the jets could be disrupted by the stellar wind.

The possible influence of magnetic fields or an inhomogeneous wind
remain to be tested. At present, we compute the evolution of Jet 2
in a decreasing density atmosphere given by the wind, and study the influence that this may have in the
position of the recollimation shock and in the growth of
instabilities.

\section*{Acknowledgments}
MP acknowledges support from a ``Juan de la Cierva'' contract of the
Spanish ``Ministerio de Ciencia y Tecnolog\'{\i}a''. MP acknowledges
support by the Spanish ``Ministerio de Educaci\'on y Ciencia'' and
the European Fund for Regional Development through grants
AYA2007-67627-C03-01 and AYA2007-67752-C03-02. The authors
acknowledge the BSC. V.B-R. acknowledges support by the Ministerio de Educaci\'on y Ciencia
(Spain) under grant AYA 2007-68034-C03-01. V.B-R. wants to thank the
Insituto Argentino de Astronom\'ia, and the Facultad de Ciencias
Astron\'omicas y Geof\'isicas de la Universidad de La Plata, for their
kind hospitality.

\end{document}